\documentstyle[12pt,psfig]{article}

\newcommand{\beq}{ \begin{eqnarray} }
\newcommand{\eeq}{ \end{eqnarray} }
\newcommand{\beqstar}{ \begin{eqnarray*} }
\newcommand{\eeqstar}{ \end{eqnarray*} }
\newcommand{\gsim}{ \mathop{}_{\textstyle \sim}^{\textstyle >} }
\newcommand{\lsim}{ \mathop{}_{\textstyle \sim}^{\textstyle <} }

\newcommand{\KEV}{ {\rm keV} }
\newcommand{\MEV}{ {\rm MeV} }
\newcommand{\GEV}{ {\rm GeV} }
\newcommand{\TEV}{ {\rm TeV} }

\begin{document}
\baselineskip 0.7cm

\begin{titlepage}

\begin{center}

\hfill KEK-TH-741\\
\hfill CERN-TH/2001-022\\
\hfill YITP-01-04\\
\hfill hep-ph/yymmdd\\
\hfill \today

  {\large The fate of the B ball}
  \vskip 0.5in {\large
    Junji~ Hisano$^{(a,b)}$,
    Mihoko M.~Nojiri$^{(c)}$ and 
    Nobuchika~ Okada$^{(a)}$}
\vskip 0.4cm 
{\it 
(a) Theory Group, KEK, Oho 1-1, Tsukuba, Ibaraki 305-0801, Japan
}
\\
{\it 
(b) Theory Devision, CERN, 1211 Geneva 23, Switzerland
}
\\
{\it 
(c) YITP, Kyoto University, Kyoto 606-8502, Japan
}
\vskip 0.5in

\abstract { The gauge-mediated SUSY-breaking (GMSB) model needs
  entropy production at a relatively low temperature in the thermal
  history of the Universe for the unwanted relics to be diluted.  This
  requires a mechanism for the baryogenesis after the entropy
  production, and the Affleck and Dine (AD) mechanism is a promising
  candidate for it.  The AD baryogenesis in the GMSB model predicts
  the existence of the baryonic Q ball, that is the B ball, and this
  may work as the dark matter in the Universe. In this article, we
  discuss the stability of the B ball in th presence of baryon-number
  violating interactions.  We find that the evaporation rate increases
  monotonically with the B-ball charge because the large field value
  inside the B ball enhances the effect of the baryon-number-violating
  operators.  While there are some difficulties to evaluate the
  evaporation rate of the B ball, we derive the evaporation time
  (lifetime) of the B ball for the mass-to-charge ratio $\omega_0\gsim
  100\;\MEV$.  The lifetime of the B ball and the distortion of the
  cosmic ray positron flux and the cosmic background radiation
  from the B ball evaporation give constraints on the baryon
  number of the B ball and the interaction, if the B ball is the dark
  matter.  We also discuss some unresolved properties of the B ball.
  }
\end{center}
\end{titlepage}
\setcounter{footnote}{0}
\section{Introduction}

The gauge-mediated SUSY-breaking (GMSB) model \cite{Giudice:1999bp} is
one of the complete models to solve the FCNC problem in the
supersymmetric (SUSY) extension of the Standard Model (SM). In a
typical GMSB model \cite{Dine:1996ag}, the dynamical SUSY-breaking
scale is of the order of $10^7\;\GEV$, so that the SUSY-breaking
masses of the order of the weak scale ($m$) are generated in the SUSY
SM.  This predicts a gravitino with a mass $m_{3/2}=100\;\KEV$, which
is beyond the closure limit of the Universe, $m_{3/2}<2h^2\;\KEV$
\cite{Pagels:1982ke}.  It is very unlikely that such a small gravitino
mass can be generated in the GMSB model, while there is an exceptional
extension for it \cite{Izawa:1997gs}. Also, the string moduli may
supply another problem in cosmology, since it is expected to have
a mass of the order of $m_{3/2}$.

This implies that there must be a substantial entropy production in
such a way that they can be diluted in the thermal history of the
Universe, we therefore have to consider a possible mechanism of
baryogenesis at a relatively low temperature. A promising candidate
for it is the Affleck and Dine (AD) mechanism
\cite{Affleck:1985fy}\cite{deGouvea:1997tn}. One of the important
predictions of the AD baryogenesis in the GMSB model is the existence
of {\it a stable Q ball} \cite{Kusenko:1998si}.  It can be a candidate
for the dark matter (DM) in the Universe.

The Q ball, a non-topological soliton, is a coherent state of a
complex scalar field \cite{Coleman:1985ki}. The existence and the
stability come from a global U(1) quantum number. In the SUSY SM the Q
ball is composed of squarks and/or sleptons with baryon ($B$) or
lepton ($L$) numbers \cite{Kusenko:1997zq}. In the GMSB model, the
flat directions $\phi$ composed of the squarks and/or sleptons are
lifted up by at most a logarithmic potential for the large field value
due to the SUSY breaking \cite{deGouvea:1997tn}. This leads to the
existence of the stable Q ball. The mass of the Q ball originated from
the flat direction is proportional to ${\bf m} Q^{\frac34}$, not to $m Q$,
because of the logarithmic potential. Here, ${\bf m}$ is the mass scale of
a logarithmic potential (${\bf m} \simeq 10^{3-5}\;\GEV$). Therefore, the
baryonic Q ball, the B ball, may be stable against the decay into
nucleons if the baryon number is sufficiently large ($Q\gsim
10^{12}$), since the mass-to-charge ratio $\omega_0$ can become less
than $1\;\GEV$.

The AD baryogenesis, which is the natural candidate for 
baryogenesis in the GMSB model,  as mentioned above, can generate such
large B balls.  In the final stage of the AD baryogenesis, the coherent
state of the AD scalar field, which consists of the flat direction
$\phi$, becomes unstable and instabilities develop.  The Q ball is
formed as a result of the fluctuation glowing.  This behaviour of the
AD field has been justified by numerical simulations \cite{qsim},
and the largest Q ball charge is proportional to the initial field
value of the AD field.  Then, the B ball dark matter is an important
prediction in the GMSB model, assuming the AD baryogenesis.
 
The B ball DM search has already given a constraint on this scenario.
Since the B ball with larger baryon number becomes stabler, the B ball
absorbs nucleons and emits an energy of about $1\;\GEV$ per a nucleon
when the B ball collides with nucleons. This process is known as the
KKST process, and it is similar to the monopole-catalyzed proton decay
\cite{Kusenko:1998vp}. From the BAKSAN, Gyrlyanda, and Kamiokande
experimental results, the constraint on the B ball charge is derived
assuming that the B ball is the DM, and it should be larger than
$10^{24}$ for ${\bf m} <1\;\TEV$ \cite{Arafune:2000yv}.  This KKST process
is suppressed by the Coulomb repulsion if the B ball has positive
electric charge. The B ball has an electron cloud around it and behaves
as a heavy atom. The interaction with matter is similar to the case of
nuclearites and the B ball charge should be larger than $10^{22-30}$,
depending on the electric charge \cite{Arafune:2000yv}.

These bounds are loosened if the supergravity
contribution to the flat direction potential is included
\cite{Kasuya:2680sc}.  For a large B ball charge, the field value
inside the B ball becomes large and the supergravity contribution
to the flat direction potential is not negligible. In this case, for a
larger B ball, its radius $R_Q$ does not increase and becomes a
constant, $\sim 10/m_{3/2}$. The geometrical cross section $\pi R_Q^2$
is smaller than when the supergravity contribution is not included.

In this article we derive another constraint on the B ball DM
scenario.  The baryon number may not be an exact symmetry in nature,
and high-energy physics may violate the baryon-number conservation,
such as in the grand unified theories (GUTs). In fact, the AD
baryogenesis needs the $(B-L)$-violating operators, which kick $\phi$
to start the rotation and to generate the baryon number.  While the
interaction with only lepton-number violation can work for the AD
baryogenesis, the baryon-number-violating operators are needed to
generate for the B ball to be generated in the AD mechanism.  In this
article, assuming the existence of the baryon-number-violating
higher-dimensional operators, we evaluate the lifetime (the
evaporation rate) of the B ball. The higher-dimensional operators,
including $\phi$, enhance the evaporation rate by the large field value
inside the B ball.  Especially, for a larger B ball, the field
value becomes larger and the evaporation rate is significantly
enhanced. As a result, the B ball does not necessarily keep the
baryon number beyond the age of Universe.

The final state in the evaporation of the B ball by the
baryon-number-violating operators depends on the quantum numbers of
the B ball, the symmetries of the operators, and $\omega_0$. If the
operator violates $(B+L)$, but not $(B-L)$, and the $B$ ball does not
have lepton numbers at all, the final states are ($e^+, \pi^-$),
($\bar{\nu}, \pi^0$), and so on for $\omega_0 \gsim 100\;\MEV$.  If
$\omega_0 \lsim 100\;\MEV$, the final state consists of only lepton
and antileptons.  Photons may be included there.  Without the
knowledge about the surface dynamics on the B ball, one cannot
estimate the evaporation rate in the case $\omega_0 \lsim 100\;\MEV$.
Since the pion emission from the B ball is not allowed kinematically,
quarks emitted by the baryon-number-violating operators are bounded to
the surface or inside of the B ball for a while, and they decay or
annihilate to the lighter particles.  In this article, we restrict
ourselves to evaluating the evaporation rate in the case $\omega_0 \gsim
100\;\MEV$; we discuss what may happen for $\omega_0 \lsim
100\;\MEV$.

If $e^+$ is in the final state, the energy is of the order of
$\omega_0$ and almost monochromatic. This may change the 
cosmic ray positron flux. The existing positron flux
measurements are above about $100\;\MEV$, which works for our case of
$\omega_0 \gsim 100\;\MEV$. Also, when $\pi^0$ is in the final state,
it may distort the cosmic background radiation. We find that both
observations give more stringent constraints on the evaporation rate
of the B ball than a comparison with the age of the Universe, assuming
that the B ball is the DM.

We organize this article as follows. In the  next section, we introduce the
Q ball following Coleman's argument, and review the profile and
the quantum numbers of the B ball originated from the flat directions
of the squarks and/or sleptons in the GMSB model. Also, we show the
profile of the Q ball configuration after including the supergravity
contribution to the scalar potential.  In Section 3, the final state
of the B ball evaporation by the baryon-number-violating operators is
discussed there for both $\omega_0 \gsim 100\;\MEV$ and $\omega_0 \lsim
100\;\MEV$. In Section 4, the evaporation rate of the B ball is presented
for $\omega_0 \gsim 100\;\MEV$ and it is compared with the age of the
Universe. In Section 5, we evaluate the fluxes of the $e^+$ and $\gamma$
emitted from the evaporation of the B ball, and give constraints on the
evaporation rate from the observations. Section 6 is devoted to
conclusions and discussion.

\section{Properties of the B ball in the GMSB model}

In this section we review the Q ball in order to fix our convention,
and summarize the properties and the profiles of the B ball in the
GMSB model.  As Coleman pointed out \cite{Coleman:1985ki}, some scalar
potential with a U$_Q$(1) symmetry predicts the Q ball to be  a
non-topological soliton.  The Lagrangian of the scalar field with the
$U_Q(1)$ charge $q$ is
\begin{eqnarray}
{\cal L} &=&  |\partial_\mu \phi|^2 -V(\phi).
\end{eqnarray}
This system has two conserved quantities, the charge $Q$ and the energy $E$,  
\begin{eqnarray}
Q &=& i \int d^3x\;q \left(
\phi^\star (\partial_t \phi)
-(\partial_t \phi^\star)  \phi
\right),
\\
E &=& \int d^3 x
\left(
|\partial_t \phi|^2 
+|\partial_i \phi|^2 
+V(\phi)
\right).
\end{eqnarray}
In this system, the non-trivial lowest-energy state with $Q$ fixed,
that is the Q ball, is derived by minimizing
\begin{eqnarray}
E_{\omega} 
&=&
 E 
+\omega 
\left\{Q - i \int d^3x q 
\left(
\phi^\star (\partial_t \phi)
-(\partial_t \phi^\star)  \phi
\right)
\right\}.
\end{eqnarray}
Here, $\omega$ is the Lagrange multiplier. The time dependence on $\phi$ of
the Q ball configuration is determined  as
\begin{eqnarray}
\phi(x) &=& \tilde{\phi}({\mbox{\boldmath $x$}}) {\rm e}^{-i q \omega t}, 
\end{eqnarray}
and $E_{\omega}$ is reduced to be 
\begin{eqnarray}
E_{\omega} 
&=&
\int d^3x 
\left((\partial_i \tilde{\phi})^2 +V_{\omega}(\tilde{\phi})\right)+\omega Q,
\end{eqnarray}
where $V_{\omega}(\tilde{\phi}) = V(\tilde{\phi}) - q^2\omega^2
\tilde{\phi}^2$.  Then, the procedure to derive the $Q$ ball
configuration is by deriving a solution of $\tilde{\phi}$ for the
equation of motion
\begin{eqnarray}
\partial_r^2 \tilde{\phi}
+\frac2 r \partial_r \tilde{\phi}
-\frac12 \frac{\partial V_{\omega}(\tilde{\phi})}{\partial \tilde{\phi}}&=&0, 
\label{eqofm}
\end{eqnarray}
with the boundary condition $\partial_r \tilde{\phi}(0) =
\tilde{\phi}(\infty)=0$, and minimizing $E_{\omega}$ for $\omega$.

We now discuss the properties and the profiles of the
B ball in the GMSB model. The large B ball originates from the flat
direction $\phi$ consisting of squarks and/or sleptons in the GMSB
model.  A typical example of the renormalisable $F$ and $D$ flat
directions is given as
\begin{equation}
\bar{u}_2^{\alpha} = \frac1{\sqrt{3}} \phi \delta^{\alpha}_1,~ 
\bar{d}_1^{\alpha} = \frac1{\sqrt{3}} \phi \delta^{\alpha}_2, ~
\bar{d}_2^{\alpha} = \frac1{\sqrt{3}} \phi \delta^{\alpha}_3. 
\label{typical}
\end{equation}
Here, $\bar{u}$ and $\bar{d}$ are singlet quarks, the upper and 
lower indices for colour and generation, respectively.  The
coefficients in the right-hand sides of Eqs.~(\ref{typical}),
$1/\sqrt{3}$, is for the canonical normalization of $\phi$.  The
renormalizable $F$ and $D$ flat directions in the SUSY SM are
summarized in Ref.~\cite{Dine:1996kz}. Here, we present only the flat
directions with the baryon numbers in Table~\ref{tab1}.
\begin{table}
  \begin{center}
    \begin{tabular}{|r||c|c|}  
      \hline
      $\phi^n$ 
      & $B$& $L$\\
      \hline\hline
      $\bar{u}\bar{d}\bar{d}$
      & $-1$ & 0  \\
      \hline
      $QQQL$
      & 1  & 1  \\
      \hline
      $\bar{u}\bar{u}\bar{d}\bar{e}$
      &$-1$  &$-1$  \\
      \hline
      $QQQQ\bar{u}$
      & 1  & 0  \\
      \hline
      $\bar{d}\bar{d}\bar{d}LL$
      &$-1$ &  2  \\
      \hline
      $\bar{u}\bar{u}\bar{u}\bar{e}\bar{e}$
      &$-1$ & $-2$  \\
      \hline
      $\bar{u}\bar{u}\bar{d}\bar{d}\bar{d}\bar{d}$
      &$-2$ & 0   \\
      \hline
      $QQQQ\bar{d}LL$
      & 1 & 2   \\
      \hline
      $QQQLLL\bar{e}$
      & 1 & 2   \\
      \hline
      $\bar{u}\bar{u}\bar{u}\bar{d}\bar{d}\bar{d}\bar{e}$
      &$-2$ &$-1$   \\
      \hline
    \end{tabular}
  \end{center}
  \caption{The renormalisable $F$ and $D$ flat directions in the SUSY SM
    with the baryon numbers.  $Q$ and $L$ stand for the doublet quarks and 
    leptons, and $\bar{u}$, $\bar{d}$, and  $\bar{e}$  for singlet quarks 
    and  leptons. }
\label{tab1}
\end{table}
Here, $Q$ and $L$ stand for the doublet quarks and leptons, and
$\bar{e}$  for singlet leptons. We suppressed the gauge and
generation indices.  The charge for the B ball composed of the flat
direction follows the charge of this direction. This is important
for determining the final states in the evaporation of the B ball.

Which type of the B ball listed in Table~\ref{tab1} is the DM in the
Universe as the result of the AD mechanism? For a fixed baryon number
$Q$, the stable B ball should be only the lightest among those of the
above flat directions. Even if the AD mechanism creates heavier B
balls with the same baryon number $Q$ as the lightest one, they change
their lepton numbers by emitting neutrinos and/or anti-neutrinos, and
transit to the lightest one. However, the transition rate should be
suppressed, and the transition time might be longer than the age of
the Universe. This is because the potential barrier between the two
states is very high due to the large field values. Therefore, we will
not calculate the transition rate to the lightest state, however, 
discuss the evaporation of the B ball generically.

Next, we show the profiles of the B ball in the GMSB model. In that
model the flat direction has a logarithmic potential due to the
radiative correction from the messenger sector \cite{deGouvea:1997tn}.
This implies that a larger B ball with a larger field value is
energetically favoured for the unit charge, which  leads to a 
stable B ball. In this article the scalar potential of the flat
direction $\phi$ in the GMSB model is simplified as
\begin{eqnarray}
V_{\rm GMSB} &=& \frac{(m M)^2}2 \left(\log\left(1+\frac{|\phi|^2}{M^2}\right)\right)^2.
\label{potential_GMSB}
\end{eqnarray}
Here, $m$ is the SUSY breaking scale in the SUSY SM ($m \sim $
$O(10^{2-3})\;\GEV$), and $M$ is for the messenger quark mass in the GMSB model. In a
typical model, the messenger quark mass is of the order of $10^5$GeV.
While this double-log potential realizes the behaviour of the exact
potential for $\phi\gg M$, derived in Ref.~\cite{deGouvea:1997tn},
it has a wrong behaviour for $\phi\ll M$. However, the Q ball properties
are determined by the behaviour of the potential for $\phi\gg M$, and 
this potential thus works for our purposes.\footnote{
  In Ref.~\cite{Kusenko:1998si} they adopt a single-log potential
  and take $M=m$ in the potential so that the behaviour of the
  potential for $\phi\ll M$ is realized.  We find that the field value
  inside the Q ball in our potential with $M=m$ is a few time larger
  than that in the single-log potential for a fixed $Q$.  } 
The profile of the large B ball derived by this potential is well
approximated to be
\begin{eqnarray}
\tilde{\phi} &=& \tilde{\phi}_0  \frac{\sin(q \omega r)}{q \omega r},
\label{prof_GMSB}
\end{eqnarray}
for $r<R_Q(\equiv\pi/q \omega)$, and $\tilde{\phi}=0$,  for $r\ge R_Q$.
The parameters in Eq.~(\ref{prof_GMSB}) and the B ball mass determined
from this profile are given by the B ball charge $Q$ as
\begin{eqnarray}
q \omega &=& (2 \pi \eta_0)^{\frac12} (m M)^{\frac12} (Q/q)^{-\frac14},
\nonumber\\
\tilde{\phi}_0 &=&  (\frac{\eta_0}{2 \pi} )^{\frac12} (m M)^{\frac12} (Q/q)^{\frac14},
\nonumber\\
R_Q &=& \left(\frac{\pi}{2 \eta_0}\right)^{\frac12} (m M)^{-\frac12} (Q/q)^{\frac14},
\nonumber\\
m_Q &=& \frac 43 (2 \pi \eta_0)^{\frac12} (m M)^{\frac12} (Q/q)^{\frac34}.
\label{gmsb_p}
\end{eqnarray}
In the following the flat directions are defined so that $q$, $\omega$
and $\omega_0$ are positive. The parameter $\eta_0$ is fitted
as
\begin{eqnarray}
\eta_0 \simeq 4.8 \log\frac{m}{q \omega} +7.4.
\end{eqnarray}
from the numerical calculation.  For $r \gsim R_Q$, where
$\tilde{\phi}\lsim {mM}/{q \omega}$, the above approximated solution
is not valid.  However, such a region does not contribute dominantly
to the properties of the B ball. In the following we call the B ball
given by Eqs.~(\ref{prof_GMSB}) and (\ref{gmsb_p}) as the GMSB B ball.

From Eqs.~(\ref{gmsb_p}), the energy per unit charge $\omega_0(\equiv
m_Q/Q)$ is 
\begin{eqnarray}
\omega_0&=& \frac43 \omega.
\label{w_w0_gmsb}
\end{eqnarray}
The physical meaning of $\omega$ is the energy for increasing or
decreasing the B ball charge by the unit charge, since
$(m_Q-m_{Q-1})=\omega$.

The field value $\tilde{\phi}_0$ is proportional to $Q^{\frac14}$, as in
Eq.~(\ref{gmsb_p}), and for larger $Q$ the supergravity contribution
to the scalar potential may be important \cite{Kasuya:2680sc}. The
scalar potential from the supergravity contribution is
\begin{eqnarray}
V_{\rm SUGRA} 
&=&
m_{3/2}^2 |\phi|^2
\left(1+K \log\frac{|\phi|^2}{M_G^2}\right),
\end{eqnarray}
where $m_{3/2}$ is the gravitino mass and $M_G$ is the reduced Planck
mass. Here we assume the minimal supergravity for simplicity, and this
potential becomes dominant when $\phi$ is larger than $m M/m_{3/2}$.
The second term in the bracket of the right-hand side comes from
one-loop correction. If it comes from the gauge interaction, $K$ is
negative and $O(10^{-2})$. The existence of the Q ball solution
requires negative $K$ \cite{Enqvist:1998si}.  In a limit where $V_{\rm
  GMSB}$ is negligible, the Q ball configuration is exactly given by
a  Gaussian form:
\begin{eqnarray}
\tilde{\phi} &=& \tilde{\phi}_0 {\rm e}^{-\frac{r^2}{2 R_Q^2}}.
\label{sugra}
\end{eqnarray}
 This solution leads to 
\begin{eqnarray}
q \omega &=& m_{3/2},
\nonumber\\
\tilde{\phi}_0 &=&  (2 \pi^{\frac32})^{-\frac12}|K|^{\frac34} m_{3/2} 
(Q/q)^{\frac12},
\nonumber\\
R_Q &=& |K|^{-\frac12} m_{3/2}^{-1},
\nonumber\\
m_Q &=& m_{3/2} (Q/q),
\label{sugra_p}
\end{eqnarray}
up to $O(K)$ \cite{Enqvist:1998si}, and then 
\begin{eqnarray}
\omega_0&=&\omega.
\end{eqnarray}
We refer to the B ball given by Eqs.~(\ref{sugra}) and (\ref{sugra_p})
as the supergravity B ball.

\begin{figure}[p]
\centerline{\psfig{file=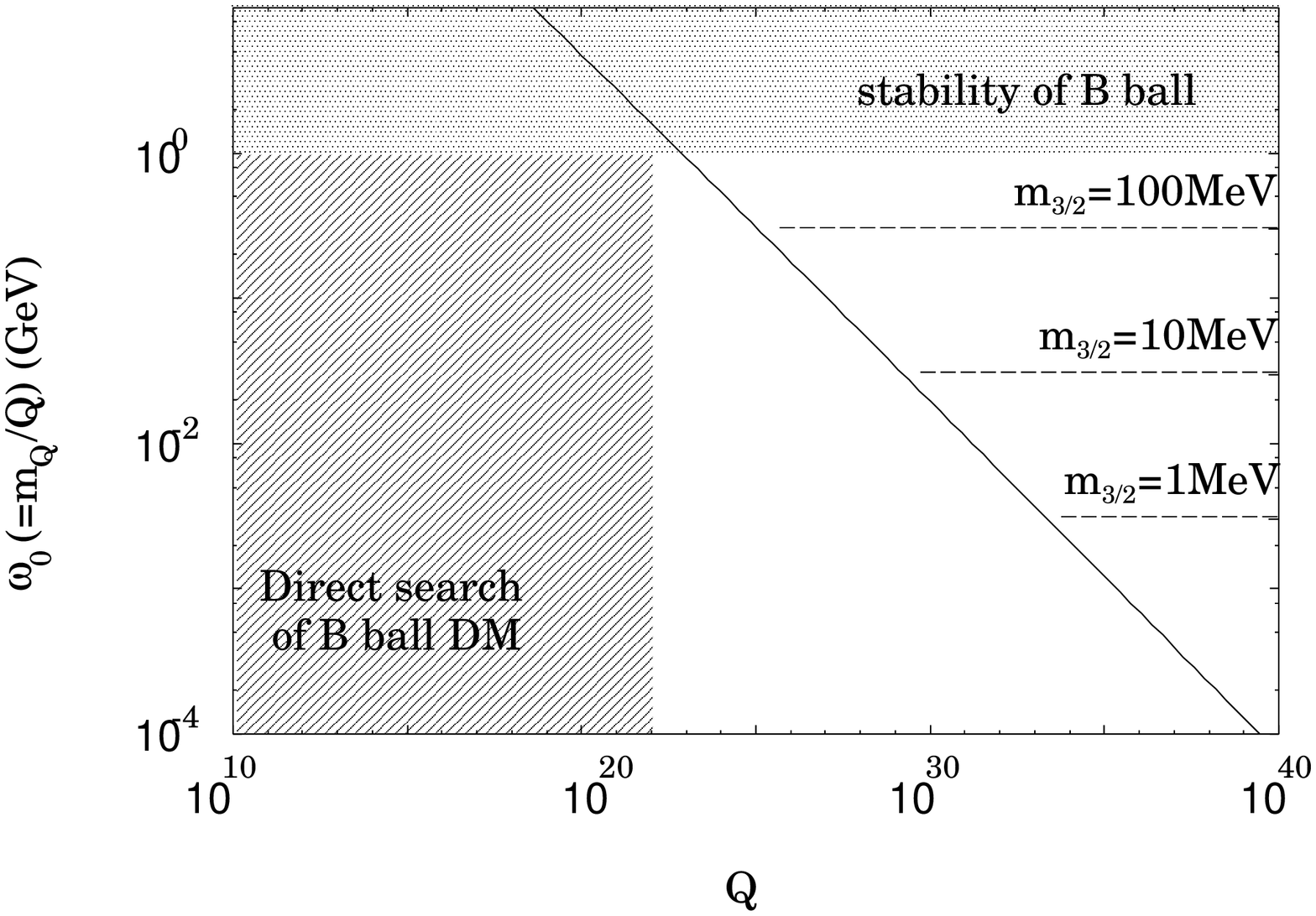,width=8cm,angle=-90}}
\caption{
  The mass-to-charge ratio of the B ball in the GMSB model. The
  horizontal axis is for the baryon number. The solid line is for the
  B ball, which comes from the flat-direction potential of the GMSB
  model. The dashed lines are for those from the supergravity
  contribution to the potential ($m_{3/2}=10^{-1}$, $10^{-2}$, and
  $10^{-3}\;\GEV$). Here, we take the SUSY breaking scale in the SUSY
  SM $m=1\;\TEV$ and the messenger scale $M=10^2\;\TEV$.  
  In the region above $1\;\GEV$ the B ball is
  unstable. The region for $Q<10^{22}$ is excluded by the direct
  search for the B ball dark matter.}
\label{fig1}
\end{figure}

In Fig.~\ref{fig1} we show the mass-to-charge ratio of the B ball,
$\omega_0$, as a function of $Q$.  The solid line is for the B ball
that comes from the flat-direction potential of the GMSB model (the
GMSB B ball).  The dashed lines are for those from the supergravity
potential (the supergravity B ball). Here, we take $m=1\;\TEV$, 
$M=10^2\;\TEV$, and
$q=1/3$, assuming that the B ball comes from $\bar{u}\bar{d}\bar{d}$.
If taking larger  $m$ or $M$, the solid line is shifted to the upper side.
For smaller $q$, both the solid and dashed lines go up.  In the region
$\omega_0> 1\;\GEV$, the B ball is unstable and decays into nucleons.
While the result of the direct search for the B ball DM depends on the
electric charge, the region for $Q\lsim 10^{22}$ is excluded for any
electric charge \cite{Arafune:2000yv}. 

Before finishing this section, we comment on the electric charge of
the B ball. This should be below the maximal electric charge
\cite{Lee:1989ag} and negligible with respect to the baryon number.
However, there is no symmetry to forbid the B ball to have an electric
charge. For example, in a flat direction of Eq.~(\ref{typical}), zero
electric charge means that the field values of $\bar{u}_1$,
$\bar{d}_1$, and $\bar{d}_2$ are equal to each other. However, this
equality might be violated by SUSY breaking and the B ball may
have a non-zero electric charge. In order to determine it, we also
need to know the details of the scalar potential.

\section{Evaporation of the B ball by the baryon-number-violating operators}

In next section we will evaluate the evaporation rate of the B ball by
the baryon-number-violating operators, using a technique described
by Cohen {\it et al.} \cite{Cohen:1986ct}. However, this technique is
not applicable to cases including scattering, annihilation, or decays
of the fermions bounded inside  the Q ball.  The energy release in
the evaporation per unit charge is $\omega$.  This means that we
cannot evaluate the evaporation rate for $\omega\lsim 100\;\MEV$, where
no mesons can be emitted from the B ball. Here, we discuss what
may happen for both cases, $\omega\gsim 100\;\MEV$ and $\omega\lsim
100\;\MEV$, while we will evaluate the evaporation rate only for 
$\omega \gsim 100\;\MEV$ in the next section.

We assume the SUSY SM with the R parity conservation.  Then the
baryon-number-violating operators are given by $F$ terms of the
effective operator with dimension larger than 5 or $D$ terms with
dimension larger than 6.  The lowest baryon-number-violating
operators in the $F$ terms are given as
\begin{eqnarray}
\cal{L} &=& \frac1{M_5} Q_i Q_j Q_k L_l|_{\theta^2}
+ 
\frac1{M_5'} \bar{u}_i \bar{e}_j \bar{u}_k \bar{d}_l|_{\theta^2} +{\rm h.c.},
\end{eqnarray}
where $i,j,k,l$ are for the generations ($i\ne k)$. These interactions
change the charges of the B ball by $\Delta(B-L)=0$ and $\Delta(B+L)=-
2$.

Let us discuss typical examples. First, we assume that the B ball is composed
of the flat direction in Eq.~(\ref{typical}),
$\bar{u}_2\bar{d}_1\bar{d}_2$.  Inserting Eq.~(\ref{typical}) in the
above operators, we get the interaction\footnote{
Notice that we define that the charge of $\phi$ is positive.
}
\begin{eqnarray}
  &\frac1{3 M_5'}  \phi^{\star 2} \bar{u}_1 \bar{e}_1.&
\label{typical_op}
\end{eqnarray}
Then, the final state is $(\pi^- e^+)$ for one baryon number. Here,
$d_1$ in the pion is supplied by the surface of the B ball.  The
production $(\pi^- e^+)$ is allowed from the kinematics if $\omega
\gsim 210\;\MEV$, since the typical momentum for each parton is
$\sim \frac13 \omega$. (This will be shown in the next section.)

Here, notice there are other interactions, such as
\begin{eqnarray}
&\frac1{3 M_5'}  \phi^{\star 2} \bar{u}_3 \bar{e}_3.&
\label{typical2}
\end{eqnarray}
Because of kinematics, the primary fermions in this interaction,
$\bar{u}_3$ and $e^+_3$, cannot be emitted from the B ball. The
primary fermions may be bounded on the surface or inside of  the B ball
for a while, and decay or annihilate into the lighter states. In fact,
$u_3$ and $e_3$ behave as a massless particle inside the Q ball if
the above operator (\ref{typical2}) is negligible.  In the
conventional SUSY-GUT, the baryon-number-violating dimension-5
operators are proportional to the fermion masses \cite{dim5}; this
process may then be enhanced and dominate over the others. If
this process dominates, the final state and the spectrum may be
different from the one mentioned above.  We need a technique to calculate
the transition rate of the primary quark, bounded in the B ball, to
lighter states in order to derive the evaporation rate of the B ball.
Keeping this possibility in mind, we will continue our discussion.

If $\omega \lsim 210\;\MEV$, all quarks are bounded inside or on the
surface of the B ball. This situation is also similar to the case
mentioned above, and we cannot evaluate the evaporation rate. The
final states are expected to be ($e^+$, $e^-$, $\nu$, $\bar{\nu}$,
$\bar{\nu}$) or ($\gamma$'s, $\nu$, $\bar{\nu}$, $\bar{\nu}$).

Next, if the B ball is composed of the flat direction $\bar{u}_1
\bar{e}_1 \bar{u}_2 \bar{d}_1$, the final states are ($\pi^0$, $e^+$,
$e^-$) or (2 $\pi^0$) for $\omega\gsim 280\;\MEV$. On the other hand,
($\pi^0$, $\bar{\nu}$, $\nu$) is also included in the final states for
the B ball composed of $Q_1 Q_1 Q_2 L_1$. The typical momentum of each
fermion is $\omega/4$ (the momentum of the pion is double of this).
These arguments are applicable to the other B balls. The exceptions
are cases where the B ball consists of
\begin{eqnarray}
\bar{u}\bar{u}\bar{u}\bar{d}\bar{d}\bar{d}\bar{e},&&
\bar{u}\bar{u}\bar{d}\bar{d}\bar{d}\bar{d}.
\label{b2}
\end{eqnarray}
In these cases, the dimension-5 operators are not effective, since the
interaction with $\Delta B=\pm2$ is needed, and the dimension-7
operators with $\Delta B=2$ may work well,
\begin{eqnarray}
{\cal{L}}=\frac{1}{M^3_7}\bar{u}_i\bar{u}_j\bar{d}_k\bar{d}_1\bar{d}_2\bar{d}_3,
\end{eqnarray}
where $i \ne j$. The final states are ($2 \pi^0$, $e^-$),
($\pi^+$, $\pi^-$, $e^-$) for the former in Eq.~(\ref{b2}), and
($2 \pi^0$), ($\pi^+$, $\pi^-$) for the latter if the
kinematics allows the processes.

Finally, we comment on the baryon-number-violating operators in the
$D$ terms.  The lowest ones are
\begin{eqnarray}
\cal{L} &=& 
\frac{1}{M_6^2}\bar{d}^{\dagger}\bar{u}^{\dagger} QL|_{\theta^2\bar{\theta}^2}
+
\frac{1}{M_6'^2}\bar{u}^{\dagger}\bar{e}^{\dagger} QQ|_{\theta^2\bar{\theta}^2}.
\end{eqnarray}
Since the relevant terms in the $D$-term operators have a derivative of
the scalar, the evaporation rate is suppressed by $\omega/M_6$,
with respect to the $F$-term operators.

\section{Evaporation rate of the B ball}

In this section we present the evaporation rate of the B ball obtained
by the baryon-number-violating operators. The technique to calculate
it was developed by Cohen {\it et al.} \cite{Cohen:1986ct}. They
evaluate the evaporation rate of the L ball to neutrinos by the
lepton-number-conserving interaction. The evaporation process to
neutrinos is equivalent to the neutrino pair production on the L ball
background.  They construct a quantum field theory preserving a
symmetry on the L ball background, a simultaneous time translation and
$L$ phase rotation, and derive the evaporation rate through the
Bogoliubov transformation between the creation and annihilation
operators in the asymptotic fields of neutrino at $t\rightarrow \pm
\infty$. The result is given by the transition amplitude to the
outgoing anti-neutrino from the incoming neutrino.

We generalize their result and apply it to the B ball evaporation.
In our case the interactions to create the fermion pairs are
baryon-number-violating.  However, we may use the formula as the
zeroth order of the Yukawa coupling constants. Also, if the
interactions are the dimension-5 operators, they preserve $(B-L)$.
Then, when the B ball is composed of $\bar{u}\bar{d}\bar{d}$, we can
use their result by regarding the B ball as the (B$-$L) ball. In this
section, we first show the properties of the Q ball evaporation
process. After that, we present the evaporation rate of the B ball and
compare it with the age of the Universe.

Since the generalization of the technique developed by Cohen is
straight-forward, we summarize the result without repeating their
calculation here.\footnote{
  Eq.~(3.4) in Ref.~\cite{Cohen:1986ct} has a typo. The second term
  in the bracket has wrong signs.  } 
The Yukawa interaction contributing to the evaporation is
\begin{eqnarray}
{\cal L} &=&  \phi \psi_1\psi_2
\end{eqnarray}
where $\phi$ is replaced by the Q ball background. The global U$_Q$(1)
charge for the scalar $\phi$ is 1, while those for fermions, $\psi_1$
and $\psi_2$, are $p$ and $(-p-1)$, respectively.  This U$_Q$(1)
symmetry  stabilizes of the Q ball.  When adopting the thin-wall
approximation for the Q ball background, $\phi$ is taken as
\begin{eqnarray}
\phi &=& \left\{
\begin{array}{cc}
v ~{\rm e}^{-i \omega t}&(r \le R_Q),\\
0&(r > R_Q).
\end{array}
\right.
\label{thin}
\end{eqnarray}
In this set-up, the evaporation rate of the Q ball is given to be
\begin{eqnarray}
\frac{dQ}{dt} 
&=&
\sum_{j=1/2}^\infty \int^{\omega}_0 \frac{d k}{4 \pi} (2j+1) |T(k,j)|^2,
\label{fromula1}
\end{eqnarray}
where $k$ and $j$ are the energy and the total orbit momentum for the
fermion. The explicit form of the transition amplitude $T(k,j)$ is
given to be
\begin{eqnarray}
T(k,j)^{-1}&=&
 \frac{w_0-k}{v k} 
\frac{(h^{(1)}_{++} j_{--} - h^{(1)}_{-+} j_{+-}) 
      (h^{(2)}_{+-} j_{-+} - h^{(2)}_{--} j_{++}) w_{-}' }
     {(h^{(1)}_{++} h^{(2)}_{-+} - h^{(1)}_{-+} h^{(2)}_{++}) 
      (j_{-+}j_{+-} - j_{--} j_{++} )}
\nonumber\\
    &-& 
 \frac{w_0-k}{v k} 
\frac{(h^{(1)}_{++} j_{-+} - h^{(1)}_{-+} j_{++})
      (h^{(2)}_{+-} j_{--} - h^{(2)}_{--} j_{+-}) w_{+}' }
    {(h^{(1)}_{++} h^{(2)}_{-+} - h^{(1)}_{-+} h^{(2)}_{++}) 
      (j_{-+}j_{+-} - j_{--} j_{++} )}.
\label{trans}
\end{eqnarray}
The transition amplitude is calculated assuming that $\psi_1$ and
$\psi_2$ are massless. Here,
\begin{eqnarray}
h^{(i)}_{\pm +} &=& h^{(i)}_{j\pm\frac12} (k R_Q) ~~~~~~~(i=1,2),\nonumber\\
h^{(i)}_{\pm -} &=& h^{(i)}_{j\pm\frac12} ((\omega-k) R_Q) ~~~~~~~(i=1,2),\nonumber\\
j_{\pm +} &=& j_{j\pm\frac12} (k_+ R_Q), \nonumber\\
j_{\pm -} &=& j_{j\pm\frac12} (k_- R_Q), \nonumber\\
w'_\pm &=& k+k_\pm-\omega,
\end{eqnarray}
with $k_\pm=\omega/2 \pm \sqrt{(k-\omega/2)^2-v^2}$. The functions,
$h^{(i)}_j(x)$ $(i=1,2)$ and $j_j(x)$, are the Hankel and Bessel
functions. The amplitude $T(k,j)$ has following properties:
\begin{eqnarray}
{\rm i)} && T(0,j)=T(\omega,j)=0, \nonumber\\
{\rm ii)} && T(k,j)=T(\omega-k,j), \nonumber\\
{\rm iii)} && \mbox{$T(k,j)$ is independent of the fermion charge $p$.}
 \nonumber
\end{eqnarray}
Also, if $R_Q$ is not so large with respect to  $\omega^{-1}$, the
contributions from the higher $j$ modes dump quickly, and the
numerical calculation is not so difficult.\footnote{
  In Ref.~\cite{Multamaki:2000an} the evaporation rate of the Q ball
  with finite $R_Q$ is calculated. }

\begin{figure}[p]
\centerline{\psfig{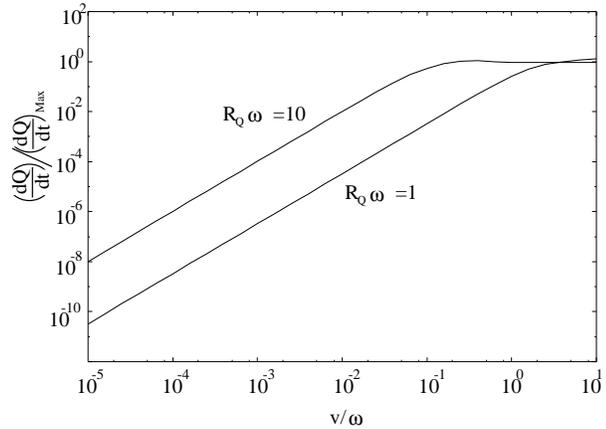}}
\caption{
  The Q ball evaporation rate for $R_Q \omega=1$ and 10 as a function
  of $v/\omega$. Here, the evaporation rate is normalized by the
  maximum value $(dQ/dt)_{\rm Max}$. (See Eq.~(\ref{max_eva}) for the
  definition.) In this calculation we adopt the thin-wall
  approximation as in Eq.~(\ref{thin}).  }
\label{fig2}
\end{figure}

In Fig.~\ref{fig2} we show the evaporation rate of the Q ball,
$dQ/dt$, for $R_Q \omega=1$ and 10 as a function of $v/\omega$. Here,
the evaporation rate is normalized by the maximum value,
\begin{eqnarray}
\left(\frac{dQ}{dt}\right)_{\rm Max} 
&=&
\frac{\omega^3R_Q^2}{48 \pi}.
\label{max_eva}
\end{eqnarray}
The qualitative behaviour of the evaporation rate is easy to 
understand. When $v/\omega>1$, the fermion pairs are produced only on
the surface of the Q ball, since the production inside the Q ball is
suppressed by the Pauli blocking. Then, the evaporation rate is
bounded by the maximum outgoing flux of the fermion pair with the
total energy $\omega$, and the maximum evaporation rate is derived
as in Eq.~(\ref{max_eva}) \cite{Cohen:1986ct}.

On the other hand, if $v/\omega<1$, the fermion pair production from
the outer shell of the Q ball, whose width is $\sim 1/v$, contributes
to the evaporation. This is because the penetration length of the
fermion inside  the Q ball, $\sim 1/v$, is larger than $1/\omega$.  Roughly
speaking, regarding the decay rate of a quantum with unit charge 
$\omega$, the evaporation rate is $\sim \omega\times(\omega v^2)\times
(R_Q^2 /v)= v \omega^2 R_Q^2$, and it is suppressed by $v/\omega$
compared with $(dQ/dt)_{\rm Max}$.  Here, $\omega v^2$ is the charge
density inside the Q ball.  This can be proved explicitly in the
limit of a large $R_Q$ \cite{Cohen:1986ct}. In this limit, the
evaporation rate becomes
\begin{eqnarray}
\frac{dQ}{dt}
&=&
\frac{\omega^2 v R_Q^2}{16}.
\end{eqnarray}
This behaviour is not valid for $1/v\gsim R_Q$. In this case, the
whole region inside the Q ball contributes to the evaporation, and
then the evaporation rate behaves, as $\sim v^2 \omega^2 R_Q^3$ as in
Fig.~\ref{fig2}.

\begin{figure}[p]
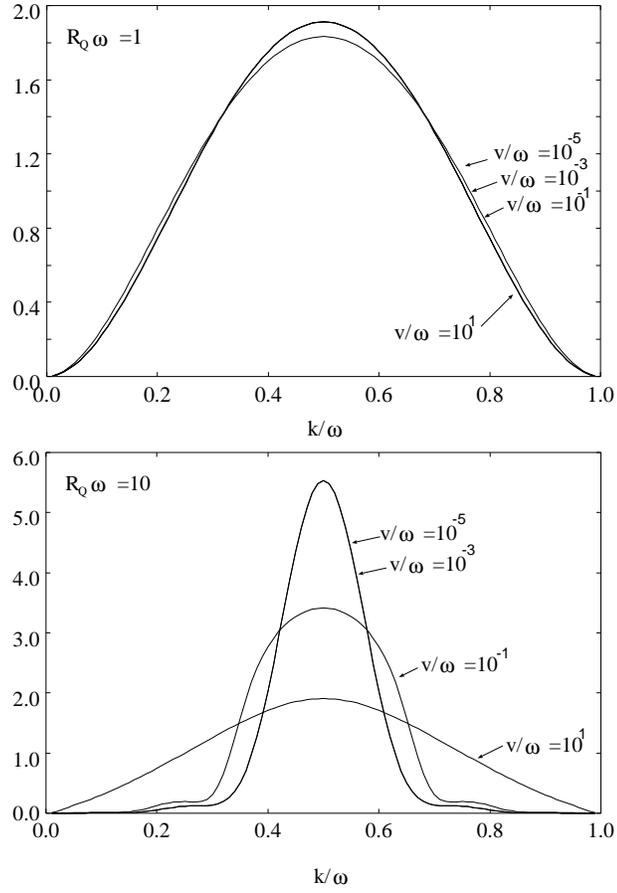

\centerline{\psfig{file=fig3_2.eps,width=8cm,angle=-90}}
\centerline{\psfig{file=fig3_1.eps,width=8cm,angle=-90}}
\caption{
  The energy spectrum of the fermion in the evaporation when
  $R_Q\omega=1$ and 10. Here we take $v/\omega=10$, $10^{-1}$,
  $10^{-3}$, $10^{-5}$. In this calculation we adopt the thin-wall
  approximation as in Eq.~(\ref{thin}).  }
\label{fig3}
\end{figure}

The above qualitative behaviour can be seen in the energy spectrum of
the fermion in the evaporation.  In Fig.~\ref{fig3} we show the
fermion energy spectrum for $v/\omega=10$, $10^{-1}$, $10^{-3}$,
$10^{-5}$, the in cases $R_Q\omega=1$ and 10.  The spectra for
$v/\omega=10$ are almost independent of $R_Q\omega$. On the other
hand, the spectrum has a peak around $\omega/2$, and it becomes
steeper for smaller $v$ and larger $R_Q$.

Now we have prepared for calculating the evaporation rate of the B
ball. Here, we use the parametrizations given in
Eqs.~(\ref{gmsb_p}) and (\ref{sugra_p}) for the thin-wall approximation.
This approximation may make an $O(1)$ error for the evaluation of the
evaporation rate. 

In order to make our discussion clear, we assume that the B ball is
composed of the flat direction $\bar{u}_2\bar{d}_1\bar{d}_2$
(Eq.~(\ref{typical})), and that evaporates into $(\pi^- e^+)$ by
the interaction in Eq.~(\ref{typical_op}).  In this case, the $\phi$,
which is canonically normalized, has the baryon number $q=1/3$, and
Eq.~(\ref{typical_op}) becomes
\begin{eqnarray}
{\cal L} &=&  \frac{\tilde{\phi}^2}{3 M_5'} {\rm e}^{i \frac23 \omega t} 
\bar{u}_1 \bar{e}_1
\label{typical_op1}
\end{eqnarray}
inside the B ball. We can apply the above formula by inserting 
\begin{eqnarray}
v &=& \frac{\tilde{\phi}^2}{3 M_5'} 
\end{eqnarray}
and rescaling $\omega$ due to the charge definition.  As mentioned in
the last section, $e_1^+$ and $\bar{u}_1$ in the final state share $2
\omega/3$ from this interaction.  The $d_1$ in $\pi^-$ is supplied
from the surface, since $d_1$ is heavy with the gaugino inside the B
ball. The $d_1$ shares $2 \omega/3$ with the other $d_1$ in the final
state. The time scale of the $d_1$ emission is of the order of $3
\omega^{-1}/2$, and the effect on the evaporation rate is
negligible.\footnote{
  The emission of $d_1$ is similar to the case of $v/\omega>1$ in
  Fig.~\ref{fig2}; the emission rate is understood from the
  analogy.  }
Here we have to note that $\pi^-$ is a pseudoscalar, and a chirality
flip is required to create a $\pi^-$ from $d_1$ and $\bar{u}_1$.  The
associated suppression $F$ may be of the order of $(m_q/f_\pi)^2\sim
10^{-3}$, where $f_\pi$ is the pion decay constant, because it comes
from the pion current interaction $J^\mu \partial_\mu\pi/f_\pi$. In
this article we do not attempt to estimate this suppression factor
precisely. We take $F=1$ when we present our numerical result.

\begin{figure}[p]
\centerline{\psfig{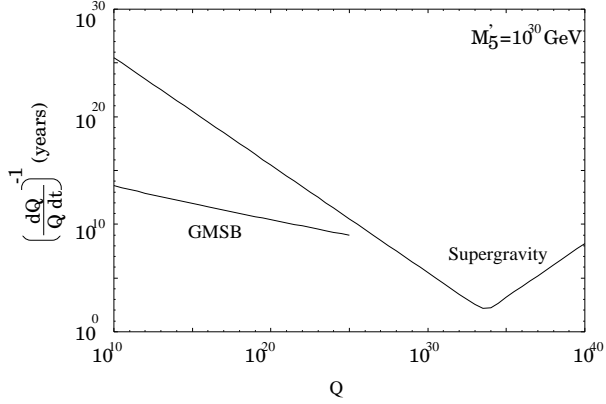}}
\caption{
  The evaporation time $(dQ/Qdt)^{-1}$ of the B ball composed of
  $\bar{u}_2\bar{d}_1\bar{d}_2$.  The two solid lines are for the
  cases the B ball originates from the GMSB or supergravity scalar
  potentials.  We take $m=1\;\TEV$ and $M=10^2\;\TEV$ for the GMSB B ball line and
  $m_{3/2}=300\;\MEV$ for the supergravity B ball line. For the other
  parameters for the B ball configuration, see the text. We plot the
  lines for $\omega>210\;\MEV$.  The suppression factor of the
  dimension-5 operator $M_5'$ is taken to be $10^{30}\;\GEV$.}
\label{fig4}
\end{figure}

In Fig.~\ref{fig4} we show the evaporation time $(dQ/Qdt)^{-1}$ of the
B ball composed of $\bar{u}_2\bar{d}_1\bar{d}_2$.  The two solid lines
are for the cases the B ball originates from the GMSB or supergravity
scalar potentials.  Here, for the supergravity B ball, we fix $m_{3/2}
= 300\;\MEV$ and $|K|^{-\frac12}=10$ and use Eq.~(\ref{sugra_p}) for
$\phi$, $R_Q$, and $\omega$. Also, for the GMSB B ball, we use
Eq.~(\ref{gmsb_p}) with $m=1\;\TEV$ and $M=10^2\;\TEV$. 
We plot the line of the GMSB B
ball for $\omega>210\;\MEV$. The suppression factor of the
dimension-5 operator $M_5'$ is taken to be $10^{30}\;\GEV$, which
is completely beyond the constraint from the negative search in
proton decay \cite{Goto:1999qg}.

In this figure, the evaporation time of the supergravity B ball
decreases as $Q$ increases for $Q\lsim 10^{34}$. This is because
$v$ is much smaller than $\omega$ and $R_Q^{-1}$, and $v$ is
proportional to $Q$.  On the other hand, when $Q\gsim 10^{34}$, the
evaporation time increases since $(dQ/dt)$ is independent of $v$
for $v/\omega>1$.  If the evaporation time is less than the age of the
Universe ($t_0\simeq 10^{10}$years), the B ball cannot survive to this 
day. In this figure we take $m_{3/2} = 300\;\MEV$.  For smaller
$m_{3/2}$,  beyond the validity of our calculation, if the qualitative
nature does not change drastically, $(dQ/Qdt)^{-1}$ may become larger
by $1/m_{3/2}^3$ when $Q\lsim 10^{33}(m_{3/2}/1\;{\rm GeV})^{-1}$
and $1/m_{3/2}$  when $Q\gsim 10^{33}(m_{3/2}/1\;{\rm GeV})^{-1}$.
The evaporation time of the GMSB B ball in this figure is proportional
to $Q^{-\frac14}$. This is because $v\ll\omega$ and $v$ is
proportional to $Q^{\frac12}$, not $Q$ as the supergravity B ball.

\begin{figure}[p]
\centerline{\psfig{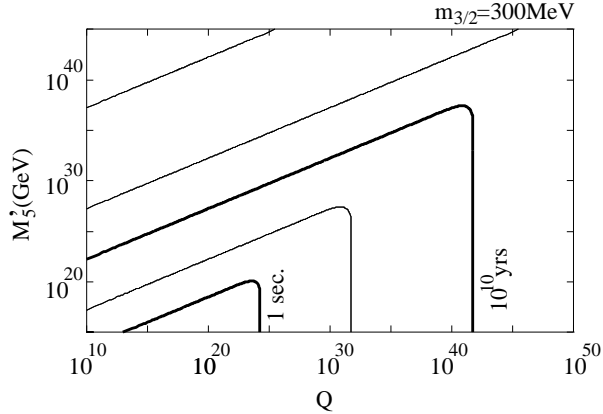}}
\caption{
  The evaporation time $(dQ/Qdt)^{-1}$ of the B ball composed
  of $\bar{u}_2\bar{d}_1\bar{d}_2$ as a function of $Q$ and $M_5'$, in
  the case where the B ball originates from the supergravity scalar
  potential with $m_{3/2}=300\;\MEV$ and $|K|^{-\frac12}=10$.  The
  thin lines are for $(dQ/Qdt)^{-1}=10^{0}$, $10^{20}$, $10^{40}$
  years, and the bold lines are for $(dQ/Qdt)^{-1}=10^{10}$ years and
  1 second.  }
\label{fig5}
\end{figure}

In Fig.~\ref{fig5} we show the evaporation time of the B ball composed
of $\bar{u}_2\bar{d}_1\bar{d}_2$ as a function of $Q$ and $M_5'$, in a
case where the B ball originates from the supergravity scalar
potential, with $m_{3/2}=300\;\MEV$ and $|K|^{-\frac12}=10$.  The thin
lines are for $(dQ/Qdt)^{-1}=10^{0}$, $10^{20}$, $10^{40}$ years.  The
bold lines are for the age of the Universe ($10^{10}$ years) and the
big-bang nucleosynthesis time (1 second).  For $Q\gsim 10^{38-40}$,
$\tilde{\phi}_0$ is larger than the Planck mass, and the region may be
disfavoured from the theoretical point of view. The region for $Q
\lsim 10^{23}$ is an unphysical region, since the B ball should be an
unstable GMSB B ball from Fig.~\ref{fig1}. This means that  $M_5'$ should 
be larger than $10^{28}\;\GEV$ at least so that the B ball behaves as 
the DM of the Universe when $m_{3/2}=300\;\MEV$.

If the B ball evaporates before the time of the big-bang
nucleosynthesis, it has no effect on the cosmology, except for the
dilution of the baryon number of the Universe. On the other hand, if
the B ball evaporation occurs after the nucleosynthesis and the
abundance is not negligible, the energetic particles in the final
state may destroy the success of the big-bang nucleosynthesis. Also, the
entropy production may change the expansion rate of the Universe. In
this article, we do not discuss the AD baryogenesis in detail;
however, the evaporation of the B ball by the baryon-number-violating
operators may give a constraint in the AD baryogenesis on the GMSB
model, which predicts a B ball with large $Q$.

Finally, we comment on the other B balls. If the B ball is composed of
$\bar{u}_1\bar{d}_1\bar{d}_2$, the fermions $\bar{u}_1$, $\bar{d}_1$,
and $\bar{d}_2$ are massive inside the B ball, and this might be in
conflict with the assumption in the above formula, Eq.~(\ref{trans}).
However, the fermion associated with the flat direction, which is a
linear combination of $\bar{u}_1$, $\bar{d}_1$ and $\bar{d}_2$, is
still massless on the flat direction condensation, and the above
formula is thus still applicable.

\section{Constraint from observation of the cosmic rays}

Even if some relic particle has a lifetime longer
than the age of the Universe, the decay products may distort the
cosmic-ray background.  This leads to a  constraint on the decaying DM. In
fact, the relic axion \cite{Turner:1987tb} and the relic Kaluza-Klein
graviton \cite{Hall:1999mk} are constrained from the the cosmic
diffused gamma ray. In this section we derive constraints on the B
ball evaporation rate from the cosmic diffused gamma background and the
cosmic ray position flux.

First, we start from the position flux induced by the B ball
evaporation.  The positrons in the evaporation of the B ball have an
almost monochromatic energy spectrum if the positron is the primary
fermion in the evaporation process and if $R_Q \omega \gsim 1$ and
$v/\omega\ll 1$, as discussed in the previous section.  However, the
positrons are diffused by the galactic magnetic field and lose their
energy through the inverse Compton and the synchrotron processes, by
the starlight and the cosmic microwave background. In this article, we
consider the standard diffusion model for the propagation of positrons
in the galaxy, which was summarized in Ref.~\cite{Baltz:1999xv}.  In
that article, the positron flux from the neutralino annihilation in
the Halo was calculated, assuming that the neutralino is the DM in the
Universe.

In their formulae, the diffusion zone of the positron in our galaxy is
a slab of the thickness $2L\simeq6$kpc, and the positron density 
becomes zero outside that region, since the positrons escape freely 
there. The standard diffusion-loss equation for the positron density 
spectrum $({dn_{e^+}}/{dE})$ is 
\begin{eqnarray}
\frac{\partial}{\partial t}
\frac{dn_{e^+}}{d\epsilon}
&=&
\overrightarrow{\nabla}\cdot
\left[
K(\epsilon,{\bf x})
\overrightarrow{\nabla}
\frac{dn_{e^+}}{d\epsilon}
\right]
+
\frac{\partial}{\partial \epsilon}
\left[b(\epsilon,{\bf x})
\frac{dn_{e^+}}{d\epsilon}
\right]
+
\frac{dn^{(0)}_{e^+}}{dtd\epsilon},
\end{eqnarray}
where $\epsilon=E/(1\;\GEV)$.  Here, $K(\epsilon,{\bf x})$ is the
diffusion constant, $b(\epsilon,{\bf x})$ the positron energy loss
rate, and $({dn^{(0)}_{e^+}}/{dtd\epsilon})$ the source term.

In the diffusion zone the diffusion constant $K(\epsilon)$ is
\begin{eqnarray}
K(\epsilon) &=& K_0(C+\epsilon^\alpha)
\nonumber\\
&\simeq& 3\times10^{27} 
\left(3^{0.6}+{\epsilon}^{0.6}\right)\; {\rm cm}^2\;{\rm s}^{-1}
\end{eqnarray}
for $E \lsim 3\;\GEV$, and the positron energy loss rate $b(\epsilon)$
is
\begin{eqnarray}
b(\epsilon)&=&\frac{1}{\tau_E} \epsilon^2
\nonumber\\
&\simeq& 
10^{-16}
{\epsilon}^2\; {\rm s}^{-1}.
\end{eqnarray}
By deriving the stable solution for the diffusion-loss equation in the
above environment, the positron spectrum originated from the DM is
given as
\begin{eqnarray}
\frac{dn_{e^+}}{d\epsilon} 
&=&
\epsilon^{-2}
\int_\epsilon^{\infty} d\epsilon' 
\frac{dn^{(0)}_{e^+}(\epsilon')}{dtd\epsilon'} 
\tau_D(\epsilon, \epsilon').
\end{eqnarray}
Here, the energy-dependent diffusion time $\tau_D(\epsilon,\epsilon')$ is
\begin{eqnarray}
\tau_D(\epsilon,\epsilon')
&=&
\frac{1}{4 K_0 \Delta v}\sum_{n=-\infty}^{+\infty} \sum_{\pm}
{\rm Erf}\left(\frac{(-)^n L\pm 2 n L \mp z}{\sqrt{4 K_0 \tau_E \Delta v}}\right)
\nonumber\\
&& \times
\int_0^\infty dr' r' f(r') 
\tilde{I}_0\left(\frac{2 r r'}{4 K_0 \tau_E \Delta v}\right)
{\rm e}^{-\frac{(r-r')^2}{4 K_0 \tau_E \Delta v}}
\end{eqnarray}
where $r$ and $z$ are cylindric coordinates for the position of the
solar system in our galaxy ($r=8.5\;{\rm kpc}$ and $z=0$).
The function $\tilde{I}_0(x)$ is $I_0(x) {\rm e}^{-x}$ with $I_0(x)$
the modified Bessel function, and ${\rm Erf}(x)$ is the error
function.  We neglect the energy-dependent part of the diffusion
constant, and
\begin{eqnarray}
\Delta v \equiv C\left(\frac{1}{\epsilon}- \frac{1}{\epsilon'}\right).
\end{eqnarray}
The function $f(r)$ is
\begin{eqnarray}
f(r)&\equiv&
\int^L_{-L} dz= g_{\rm DM}^m(\sqrt{r^2+z^2}),
\end{eqnarray}
with $g_{\rm DM}(r_{sph})$ $(r_{sph}^2=r^2+z^2)$ the DM density profile.
The exponent $m$ is 1 for the decay process of the DM, and 2 for the
annihilation process.  In this article, we use the isothermal
distribution for the DM density profiles,
\begin{eqnarray}
g_{\rm DM}(r_{sph})&=& \frac{a^2}{r_{sph}^2+a^2},
\end{eqnarray}
with $a=5$ kpc. The $N$-body simulation suggests the cuspy density
profile at the centre of the galaxy \cite{cuspy}, and it may not be 
consistent with the isothermal distribution. 
However, the decay process is not sensitive to the density profile,
and we thus use the isothermal distribution for the DM here.

\begin{figure}[p]
\centerline{\psfig{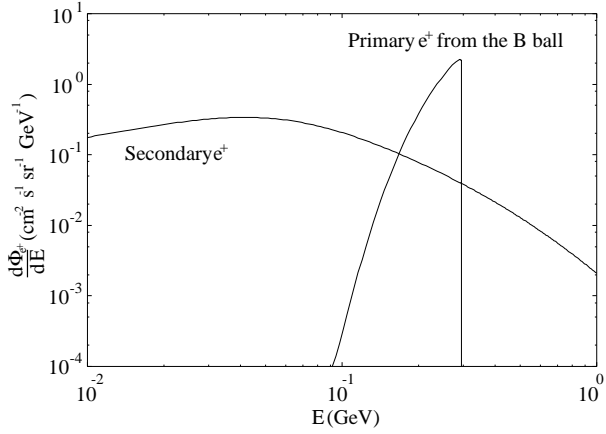}}
\caption{
  The primary and secondary positron flux spectra $d\Phi_{e^+}/dE$.
  The primary flux comes from the evaporation of the B ball, assuming
  $q \omega=300\;\MEV$, $q=1/3$, $\omega=\omega_0$, as for the
  supergravity B ball, and an evaporation time $(dQ/Qdt)^{-1}=10^{18}$
  years.  The number density is fixed by $Q \omega_0 n_0
  =0.3\;\GEV\;{\rm cm}^{-3}$. The line for the secondary positron flux
  is given in Ref.~\cite{Baltz:1999xv}.}
\label{fig6}
\end{figure}

If the positron flux comes from the evaporation of the B ball
by the baryon-number-violating dimension-5 operators,
the source term $({dn^{(0)}_{e^+}}/{dtd E})$ is 
\begin{eqnarray}
\frac{dn^{(0)}_{e^+}}{dtdE}
&=& n_0 \frac{dQ}{dt} \delta(E - q \omega).
\end{eqnarray}
Here, we simplify the spectrum of the positron as a monochromatic one.
This is justified for $R_Q \omega>1$ and $v\ll \omega$ giving a long
evaporation time, as shown in Fig.~\ref{fig3}. We take the number
density of the B ball $n_0$ as $Q \omega_0 n_0 =0.3\;{\rm GeV}\; {\rm
  cm}^{-3}$. The diffusion and the energy reduction in the propagation
makes a tail below $q \omega$ in the positron spectrum. In
Fig.~\ref{fig6} we show the primary positron flux spectrum
$(d\Phi^{(p)}_{e^+}/dE)$, which is given as $(d\Phi^{(p)}_{e^+}/dE)
\equiv (c/4\pi) ({dn_{e^+}}/{dE})$, assuming $q \omega=300\;\MEV$,
$q=1/3$, and $(dQ/Qdt)^{-1}=10^{18}\;{\rm years}$. Also, we take
$\omega_0=\omega$ as for the supergravity B ball. If
$\omega_0=4\omega/3$ as for the GMSB B ball, the primary flux is
reduced by $3/4$.

While the cosmic ray positron flux is measured for an energy larger
than about $70\;\MEV$ \cite{pf}, the theoretical estimate of the
background has uncertainties. The secondary positron flux, which comes
from the nuclear interaction of the primary cosmic rays in the
interstellar space, is estimated in Ref.~\cite{MoskStrong98}. The
result is fitted in Ref.~\cite{Baltz:1999xv} as
\begin{eqnarray}
  \frac{d\Phi_{e^+}^{(s)}}{d E} & = &  \frac{4.5 \epsilon^{0.7}}
  {1+650\epsilon^{2.3}+1500\epsilon^{4.2}}
  \;{\rm cm}^{-2}\;{\rm s}^{-1}\;{\rm sr}^{-1} \; {\rm GeV}^{-1}.
\label{sf}
\end{eqnarray}
The qualitative behaviour of the positron fraction $(e^+/(e^++e^-))$
in the cosmic ray, increasing of the positron flux at lower energy, is
realized by assuming that the positron is of a secondary origin.
However, the effect of the solar modulation from the solar wind and
the magnetosphere to both the positron and electron fluxes is larger
at lower energy. Especially, the introduction of the charge-dependent
solar modulation makes the fit to the observation worse
\cite{Baltz:1999xv}.  Therefore, we derive a conservative constraint
on the evaporation time of the B ball by imposing a condition that the
peak of the primary flux from the B ball be smaller than ten
times of the secondary flux, Eq.~(\ref{sf}), and we obtain
\begin{eqnarray}
\left(\frac{d Q}{Q dt}\right)^{-1} \gsim 2\;q\times 10^{19} {\;\rm years}
\left(\frac{Q \omega_0 n_0}{0.3\;{\rm GeV}\;{\rm cm^{-3}}}\right)
\left(\frac{q \omega}{100{\rm MeV}}\right)^{-1},
\end{eqnarray}
for $q\omega\sim 100\;\MEV$. From Fig.~\ref{fig5}, this constraint
means that $M_5'$ should be larger than $10^{32}\;\GEV$ for $Q\gsim
10^{22}$, assuming the B ball is the DM of the Universe and
$m_{3/2}=300\;\MEV$.

\begin{figure}[p]
\centerline{\psfig{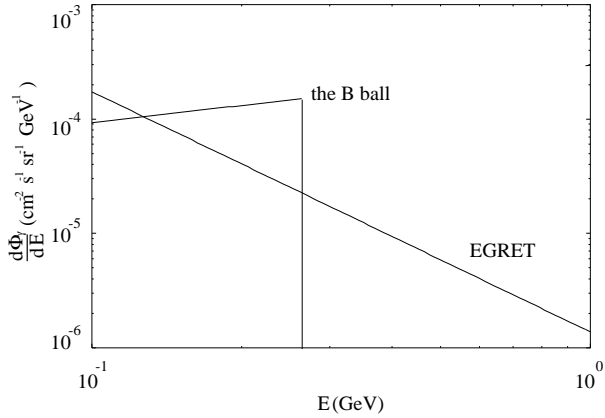}}
\caption{
  The spectrum of the gamma flux from the B ball evaporation
  $(d\Phi_{\gamma}/dE)$, assuming $q \omega=250\;\MEV$, $q=1/4$,
  $\omega=\omega_0$, and $(dQ/Qdt)^{-1}=10^{19}\;{\rm years}$.  The
  number density is fixed as $Q \omega_0 n_0 = \rho_C$ with $h=0.7$.
  The result from EGRET is also shown.}
\label{fig7}
\end{figure}

Next, let us consider the distortion of the cosmic background
radiation by $\pi^0$ from the B ball evaporation. Although the gamma,
which comes from $\pi^0\rightarrow 2 \gamma$, is almost monotonic at
production time, the energy is reduced by the red-shift.  As a result,
the energy spectrum of the gamma from the B ball evaporation is
\begin{eqnarray}
\frac{d n_{\gamma}}{d E}
&=&
3 n_0 \frac{dQ}{dt} t_0 E^{\frac12} (q\omega) ^{-\frac32},
\label{44}
\end{eqnarray}
for $E<q\omega$, assuming $(dQ/Qdt)^{-1}\gg t_0$. Here, we assume that
the B ball evaporates through the baryon-number-violating dimension-5
operators and that the $\pi^0$ has an energy $2 q\omega$.  From this
equation (\ref{44}), the flux of gamma rays for $E<q \omega$ is
\begin{eqnarray}
\frac{d\Phi_{\gamma}}{d E} & = & 
7.9\;h^2 \times10^6 \; \;{\rm cm}^{-2}\;{\rm s}^{-1}\;{\rm sr}^{-1} \;{\rm GeV}^{-1}
\nonumber\\
&\times&
\left(\frac{t_0}{(dQ/Qdt)^{-1}}\right)
\left(\frac{Q \omega_0 n_0}{\rho_C}\right)
\left(\frac{\omega_0}{0.1\; {\rm GeV}}\right)^{-1}
\left(\frac{q \omega}{0.1\; {\rm GeV}}\right)^{-\frac32}
\left(\frac{E}{0.1\; {\rm GeV}}\right)^{\frac12}.
\end{eqnarray}
Here, $\rho_C$ is the critical density of the Universe ($\rho_C=1.1
h^2\times 10^{-5}\; {\rm GeV}\;{\rm cm}^{-3}$).  In Fig.~\ref{fig7}
we show the spectrum of the gamma rays flux from the B ball evaporation
$(d\Phi_{\gamma}/dE)$, assuming $q \omega=250\;\MEV$, $q=1/4$, and
$10^{19}\;{\rm years}$ for the evaporation time $(dQ/Qdt)^{-1}$.  Here
we fix $h=0.7$ and $t_0=10^{10}\;{\rm years}$.  Also, we take
$\omega_0=\omega$ as for the supergravity B ball.

EGRET determines the extragalactic gamma ray background spectrum
between 0.1 and $\sim 50\;\GEV$ \cite{egret} as
\begin{eqnarray}
\frac{d\Phi_{\gamma}}{d E}
&=&
(7.32\pm 0.34)\times 10^{-6}
\left(\frac{E}{0.451{\rm GeV}}\right)^{-2.10\pm0.03}
\;{\rm cm}^{-2}\;{\rm s}^{-1}\;{\rm sr}^{-1}\; {\rm GeV}^{-1}.
\label{EGRET}
\end{eqnarray}
Here, imposing the condition that the peak of the gamma spectrum from
the B ball be smaller than Eq.~(\ref{EGRET}) leads us to a constraint
on the evaporation time of the B ball:
\begin{eqnarray}
\left(\frac{d Q}{Q dt}\right)^{-1} \gsim 5 \;q h^2 \times 10^{20} 
\;{\rm years}\;
\left(\frac{t_0}{10^{10}\;{\rm years}}\right)
\left(\frac{Q \omega_0 n_0}{\rho_C}\right)
\left(\frac{q \omega}{100{\rm MeV}}\right)^{0.1},
\end{eqnarray}
for $q\omega\sim 100\;\MEV$.  This is one order of magnitude stronger
than the constraint from the positron flux.

\section{Conclusions and discussion}

In this article we discuss the stability of the B ball in the GMSB
model. The B ball is predicted to exist in the GMSB model, assuming AD
baryogenesis. While the stability of the B ball comes from
the baryon-number conservation, the baryon-number-violating
interaction is required to make the baryonic AD condensation, which is
a seed for the B ball. The B ball could therefore be unstable.  We
find that a larger B ball evaporates faster, since the field value inside
the B ball enhances the evaporation rate.

The evaluation of the evaporation rate of the B ball suffers from
various difficulties, and we therefore restrict our calculation to a B
ball of the mass-to-charge ratio $\omega_0\gsim 100\;\MEV$.  We derive
the constraints on the B ball charge and the interactions, when the B
ball is composed of $\bar{u}_2\bar{d}_1\bar{d}_2$, as an example.
Neglecting the chiral suppression from the final state, the
suppression factor of the baryon-number-violating dimension-5 operator
$M_5'$ should be larger than $10^{27}(Q/10^{20})^{\frac12}\;\GEV$ for
$m_{3/2}=300\;\MEV$, so that the evaporation time $(dQ/Qdt)^{-1}$ is
longer than the age of the Universe ($t_0\simeq 10^{10}\;{\rm
  years}$). The final states of the evaporation may include the almost
monochromatic positron or $\pi^0$.  If the B ball is the DM of the
Universe, the evaporation may give a contribution to the extragalactic
gamma ray background spectrum and to the cosmic ray positron flux.
From the current data, we give constraints on the evaporation time
from the primary positron flux and the gamma ray background, for
$\omega_0\sim 100\;\MEV$, of $(dQ/Qdt)^{-1}\gsim 10^{18}$ or $10^{19}$
years, which are stronger than the constraint from the age of the
Universe.

The instability of the B ball will have some impact on the cosmology
of the GMSB model.  In this article, we do not discuss detail of the
AD baryogenesis.  The existence of the B ball may be preferred, as far
as the AD field with the baryon numbers condensates in the GMSB model. If
the B ball has an evaporation time shorter than the age of the
Universe, the evaporation may supply the entropy production and the
injection of a high-energy positron after the big-bang nucleosynthesis.
Note that the B ball may explain the reionization of the Universe,
indicated by the Gunn-Peterson test \cite{Gunn-Peterson}. It is well-known
that the Universe is ionized at $z\sim 3$, from the absence of the strong
Layman-alpha scattering light.  While this may be explained by the
huge star formation at the time, it may come from the late-time decay
of an exotic relic particle, and  the B ball may work for it.

\section*{Acknowledgements}
We would like to thank S.~Asai, G.~Giudice, S.~Kasuya, M.~Kawasaki,
L.~Roszkowski, and N.~Sugiyama for useful discussions. This work was
started at the visiting program at the YITP of the Kyoto University.
This was also supported in part by the Grant-in-Aid for Scientific
Research from the Ministry of Education, Science, Sports and Culture
of Japan, on Priority Area 707 ``Supersymmetry and Unified Theory of
Elementary Particles" (J.H.) and Grant-in-Aid for Scientific Research
from the Ministry of Education (12047217, M.M.N).

\newpage
%
%
%
\newcommand{\Journal}[4]{{\sl #1} {\bf #2} {(#3)} {#4}}
\newcommand{\PL}{\sl Phys. Lett.}
\newcommand{\PR}{\sl Phys. Rev.}
\newcommand{\PRL}{\sl Phys. Rev. Lett.}
\newcommand{\NP}{\sl Nucl. Phys.}
\newcommand{\ZP}{\sl Z. Phys.}
\newcommand{\PTP}{\sl Prog. Theor. Phys.}
\newcommand{\NC}{\sl Nuovo Cimento}
\newcommand{\MPL}{\sl Mod. Phys. Lett.}
\newcommand{\PRep}{\sl Phys. Rep.}

\newpage
\clearpage
%
%
%
%
%
%
%
%
%
%
%
%
%
%
%
%
%
%
%
%
%
%
%
%
%
%

\begin{thebibliography}{99}

\bibitem{Giudice:1999bp}
For a review, see G.~F.~Giudice and R.~Rattazzi,
Phys.\ Rep.\ {\bf 322} (1999) 419
[hep-ph/9801271].

\bibitem{Dine:1996ag}
M.~Dine, A.~E.~Nelson, Y.~Nir and Y.~Shirman,
Phys.\ Rev.\ D {\bf 53} (1996) 2658
[hep-ph/9507378].

\bibitem{Pagels:1982ke}
H.~Pagels and J.~R.~Primack,
Phys.\ Rev.\ Lett.\ {\bf 48} (1982) 223;\\
T.~Moroi, H.~Murayama and M.~Yamaguchi,
Phys.\ Lett.\ {\bf B303} (1993) 289.

\bibitem{Izawa:1997gs}
K.~I.~Izawa, Y.~Nomura, K.~Tobe and T.~Yanagida,
Phys.\ Rev.\ D {\bf 56} (1997) 2886
[hep-ph/9705228].

\bibitem{Affleck:1985fy}
I.~Affleck and M.~Dine,
Nucl.\ Phys.\ {\bf B249} (1985) 361.

\bibitem{deGouvea:1997tn}
A.~de Gouvea, T.~Moroi and H.~Murayama,
Phys.\ Rev.\ D {\bf 56} (1997) 1281
[hep-ph/9701244].

\bibitem{Kusenko:1998si}
A.~Kusenko and M.~Shaposhnikov,
Phys.\ Lett.\ {\bf B418} (1998) 46
[hep-ph/9709492].

\bibitem{Coleman:1985ki}
S.~Coleman,
Nucl.\ Phys.\ {\bf B262} (1985) 263.

\bibitem{Kusenko:1997zq}
A.~Kusenko,
Phys.\ Lett.\ {\bf B405} (1997) 108
[hep-ph/9704273].

\bibitem{qsim}
K.~Enqvist and J.~McDonald,
Nucl.\ Phys.\ {\bf B570} (2000) 407
[hep-ph/9908316];\\
S.~Kasuya and M.~Kawasaki,
Phys.\ Rev.\ D {\bf 61} (2000) 041301
[hep-ph/9909509];
Phys.\ Rev.\ D {\bf 62} (2000) 023512
[hep-ph/0002285].

\bibitem{Kusenko:1998vp}
A.~Kusenko, V.~Kuzmin, M.~Shaposhnikov and P.~G.~Tinyakov,
Phys.\ Rev.\ Lett.\ {\bf 80} (1998) 3185
[hep-ph/9712212].

\bibitem{Arafune:2000yv}
J.~Arafune, T.~Yoshida, S.~Nakamura and K.~Ogure,
Phys.\ Rev.\ D {\bf 62} (2000) 105013
[hep-ph/0005103].

\bibitem{Kasuya:2680sc}
S.~Kasuya and M.~Kawasaki,
Phys.\ Rev.\ Lett.\ {\bf 85} (2000) 2677
[hep-ph/0006128].

\bibitem{Dine:1996kz}
M.~Dine, L.~Randall and S.~Thomas,
Nucl.\ Phys.\ {\bf B458} (1996) 291
[hep-ph/9507453].

\bibitem{Enqvist:1998si}
K.~Enqvist and J.~McDonald,
Phys.\ Lett.\ {\bf B425} (1998) 309
[hep-ph/9711514];
Nucl.\ Phys.\ {\bf B538} (1999) 321
[hep-ph/9803380];
Phys.\ Lett.\ {\bf B440} (1998) 59
[hep-ph/9807269].

\bibitem{Lee:1989ag}
K.~Lee, J.~A.~Stein-Schabes, R.~Watkins and L.~M.~Widrow,
Phys.\ Rev.\ D {\bf 39} (1989) 1665.

\bibitem{Cohen:1986ct}
A.~Cohen, S.~Coleman, H.~Georgi and A.~Manohar,
Nucl.\ Phys.\ {\bf B272} (1986) 301.

\bibitem{dim5}
N.~Sakai and T.~Yanagida,
Nucl.\ Phys.\ {\bf B197} (1982) 533;\\
S.~Weinberg,
Phys.\ Rev.\ D {\bf 26} (1982) 287;\\
P.~Nath, A.~H.~Chamseddine and R.~Arnowitt,
Phys.\ Rev.\ D {\bf 32} (1985) 2348;\\
J.~Hisano, H.~Murayama and T.~Yanagida,
Nucl.\ Phys.\ {\bf B402} (1993) 46
[hep-ph/9207279].

\bibitem{Multamaki:2000an}
T.~Multamaki and I.~Vilja,
Nucl.\ Phys.\ {\bf B574} (2000) 130
[hep-ph/9908446].

\bibitem{Goto:1999qg}
T.~Goto and T.~Nihei,
Phys.\ Rev.\ D {\bf 59} (1999) 115009
[hep-ph/9808255].

\bibitem{Turner:1987tb}
M.~S.~Turner,
Phys.\ Rev.\ Lett.\ {\bf 59} (1987) 2489.

\bibitem{Hall:1999mk}
L.~J.~Hall and D.~Smith,
Phys.\ Rev.\ D {\bf 60} (1999) 085008
[hep-ph/9904267].

\bibitem{Baltz:1999xv}
E.~A.~Baltz and J.~Edsjo,
Phys.\ Rev.\ D {\bf 59} (1999) 023511
[astro-ph/9808243].

\bibitem{cuspy}
J.F.~Navarro, C.S.~Frenk and S.D.M.~Whilte,
Astrophys.\ J.\ {\bf 490} (1997)  493;\\
J.~Dubinski and R.G.~Carlberg,
Astrophys.\ J.\ {\bf 378} (1991) 496;\\
S.W.~Warren {\it et al.},
Astrophys.\ J.\ {\bf 399} (1992)  405.

\bibitem{pf}
See references in  S.~Coutu {\it et al.},  
Astropart.\ Phys.\ {\bf 11} (1999)  429;\\
also, see references in \cite{Baltz:1999xv}.

\bibitem{MoskStrong98}
I.V.~Moskalenko and A.W.~Strong, 
Astrophys.\ J.\ {\bf 493} (1998)  {694} .

\bibitem{egret}
P.~Sreekumar {\it et al.},
Astrophys.\ J.\ {\bf 494} (1998)  {523};\\
also, see F.W.~Stecker and  M.H.~Salamon, 
Proc.\ 26th ICRC, Vol.\ 3, p.~313 
[astro-ph/9909157].

\bibitem{Gunn-Peterson}
J.E.~Gunn and B.A. Peterson, Astrophys.\ J.\ {\bf 142} (1965) 1633.

\end{thebibliography}
\end{document}